\documentclass[prb,twocolumn,aps,showpacs,amsmath,amssymb,floatfix]{revtex4-2}
\usepackage{url}
\usepackage{amsmath}
\usepackage{amssymb}
\usepackage{bm}
\usepackage{graphicx}
\usepackage{hyperref}
\usepackage{dcolumn}
\usepackage{bm}
\usepackage{xcolor}
\usepackage{siunitx}
\usepackage[caption=false]{subfig}
\usepackage{multirow}
\usepackage{tabularx}
\hypersetup{
	colorlinks=true,       
	linkcolor=blue,        
	citecolor=blue,        
	filecolor=magenta,     
	urlcolor=blue         
}

\graphicspath{ {images/} }
\bibpunct{[}{]}{,}{n}{}{}

\def\m{\mathbf{m}}
\def\n{\mathbf{n}}

\def\tanh{\text{tanh}}

\begin{document}
\title{Spin-transfer torque in altermagnets with magnetic textures}
\author{Hamed Vakili}
\affiliation{Department of Physics and Astronomy and Nebraska Center for Materials and Nanoscience, University of Nebraska, Lincoln, Nebraska 68588, USA}
\author{Edward Schwartz}
\affiliation{Department of Physics and Astronomy and Nebraska Center for Materials and Nanoscience, University of Nebraska, Lincoln, Nebraska 68588, USA}
\author{Alexey A. Kovalev}
\affiliation{Department of Physics and Astronomy and Nebraska Center for Materials and Nanoscience, University of Nebraska, Lincoln, Nebraska 68588, USA}
\begin{abstract}
We predict the existence of anisotropic spin-transfer torque effect in textured altermagnets. To this end, we generalize the Zhang-Li torque to incorporate the symmetry associated with prototypical $d$-wave altermagnets and identify the spin-splitter adiabatic and nonadiabatic torques. Applying our results to domain wall dynamics induced by spin-transfer torque, we find that, in certain regimes, the spin-splitter adiabatic torque can induce domain wall precession, significantly slowing down domain wall motion. The response of the domain wall also becomes anisotropic, reflecting the $d$-wave symmetry of the altermagnet.
Furthermore, we observe that the spin-splitter adiabatic torque modifies skyrmion dynamics, inducing anisotropic skyrmion Hall effect. The above phenomena can serve as a hallmark of altermagnetism in textured magnets, distinguishing it from the behavior of ordinary antiferromagnets. 
\end{abstract}

\maketitle

Recently, a new class of materials, termed altermagnets, has received substantial attention due to potential spintronics applications~\cite{PhysRevB.102.014422,mejkal2020,PhysRevX.12.031042,PhysRevX.12.040002,Guo2023}. These materials can potentially combine the advantages of antiferromagnets, such as fast dynamics, with new features such as the $d$/$g$/$i$-wave splitting appearing in the non-relativistic band structure~\cite{PhysRevX.12.040501}. For applications relying on spin-orbit torques~\cite{RevModPhys.91.035004}, these materials offer unconventional torque contributions that arise even without spin-orbit interactions due to the spin-splitter effect~\cite{Naka2019,PhysRevLett.126.127701,PhysRevLett.128.197202,PhysRevLett.129.137201}. In addition, altermagnets are predicted to exhibit the crystal anomalous Hall effect~\cite{Feng2022,PhysRevLett.133.086503} and anisotropic magnon bands with lifted degeneracy~\cite{PhysRevLett.131.256703,PhysRevLett.131.186702}. 
The textures in altermagnets also exhibit unconventional behavior, manifested in magnetization arising in a magnetic domain wall, which in turn leads to an anisotropic Walker breakdown~\cite{Gomonay2024}.

Various effects characteristic to antiferromagnetic textures can be revisited in the context of altermagnets. The spin-transfer torque of the Zhang-Li form~\cite{PhysRevLett.93.127204} appears in textured antiferromagnets, as has been established some time ago~\cite{PhysRevLett.100.226602,PhysRevB.83.054428,PhysRevLett.106.107206,PhysRevB.94.054409}. Spin-transfer torque can be used to control both skyrmion dynamics~\cite{PhysRevLett.116.147203,PhysRevB.101.024429,Zhang2016} and domain wall dynamics~\cite{PhysRevLett.110.127208,PhysRevLett.106.107206} in antiferromagnets. A peculiar feature of current-driven skyrmions in antiferromagnets is the absence of the skyrmion Hall effect~\cite{Zhang2016,PhysRevLett.116.147203}, which can be attributed to the cancellation of the topological charge of two sublattices. Fast motion of domain walls in antiferromagnets has been predicted due to the absence of Walker breakdown~\cite{PhysRevLett.117.017202,PhysRevLett.117.107201}.
Domain walls in magnetoelectric antiferromagnets~\cite{Belashchenko2016} and ferrimagnets~\cite{Kim2020} exhibit domain wall precession, associated with oscillations between the N\'{e}el and Bloch types of domain walls.

\begin{figure}
    \centering
    \includegraphics[width=0.9\linewidth]{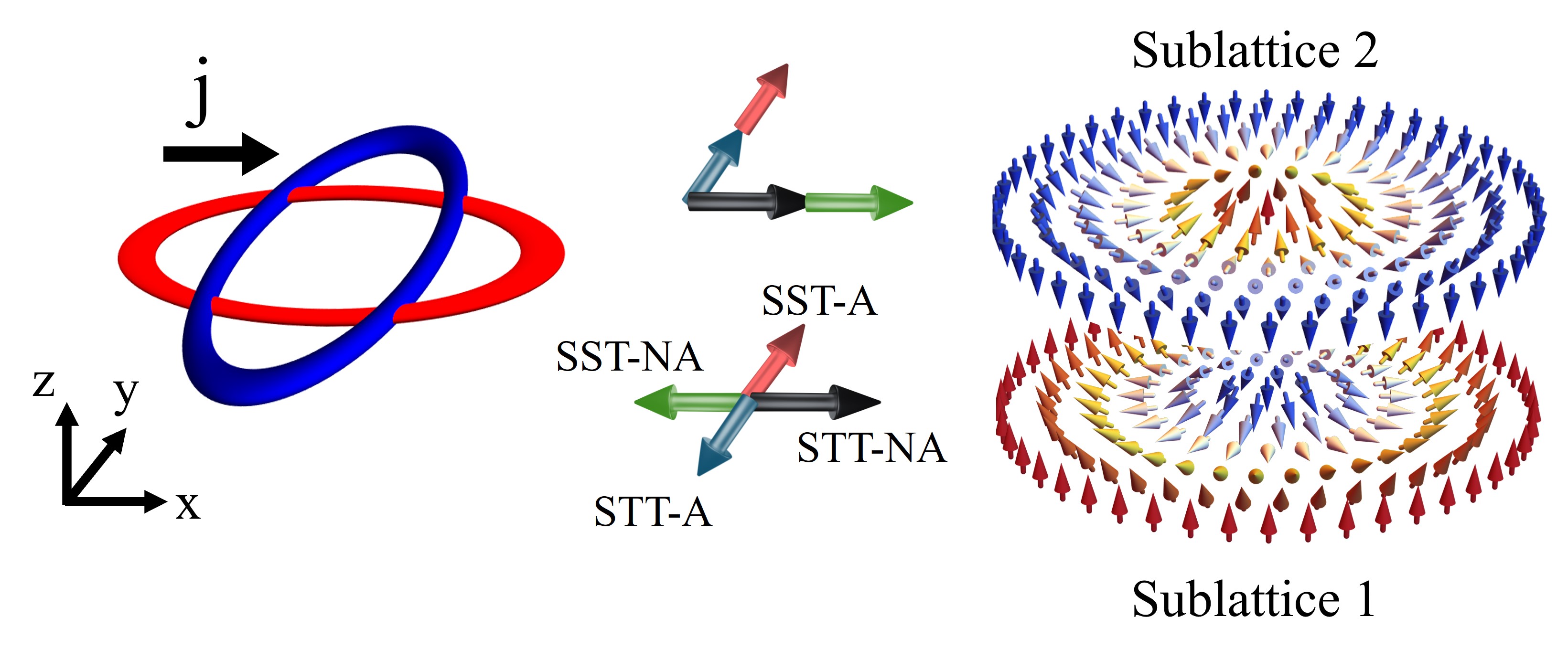}
    \caption{Schematics of the spin-transfer and spin-splitter effects in altermagnets, where the two sublattices are represented as two layers and the current direction is aligned with the altermagnetic order ($\Theta=0$). The spin-transfer effect corresponds to spin currents flowing in the same direction in both layers, while the spin-splitter effect corresponds to spin currents flowing in opposite directions in the two layers. We illustrate the directions of the forces induced on a skyrmion in the two sublattices by the adiabatic and nonadiabatic spin-transfer torques (STT-A and STT-NA) as well as the adiabatic and nonadiabatic spin-splitter torques (SST-A and SST-NA), which result in the skyrmion Hall effect.  }
    \label{fig:sst}
\end{figure}

In this Letter, we identify the spin-splitter adiabatic and nonadiabatic torques and study their effects on magnetic textures in altermagnets (see Fig.~\ref{fig:sst}). To this end, using symmetry considerations, we establish the form of the spin-splitter torques in $d$-type altermagnets and incorporate these torques into the Lagrangian formalism.  We find that the current-induced dynamics of magnetic textures is substantially modified in altermagnets. In particular, the adiabatic spin-splitter torque induces precession of the domain wall, which slows down the domain wall motion for certain directions of the charge current. For the current-induced dynamics of skyrmions, the spin-splitter adiabatic torque induces a large Magnus force, enabling the skyrmion Hall effect and much faster dynamics of skyrmions compared to the effects of conventional spin-transfer torques in atiferromagnets. 

\textit{Model and methods.} -- 
Following Ref.~\cite{Gomonay2024}, we define the Free energy of altermagnet as
\begin{align}\label{eq:hamiltonian}
    E &= d \times\int \left[H_{ex} N_s \m^2/2 +A(\partial_\alpha {\bf\n})^2 \right.\\&+\left. B(\partial_x \m \cdot \partial_x {\bf\n} - \partial_y \m \cdot \partial_y {\bf\n}) - H_{an} N_s n_z^2\right]d^2r , \nonumber
\end{align}
where ${\bf n}=({\bf m}_1-{\bf m}_2)/2$ is the N\'{e}el field and ${\bf m}=({\bf m}_1+{\bf m}_2)/2$ is the magnetization, both defined in terms of the sublattice magnetizations, $d$ is the thickness of a quasi-two-dimensional film, the antiferromagnetic exchange and anisotropy fields are described by $H_{ex}$ and $H_{an}$ respectively, $A$ describes the strength of antiferromagnetic exchange stiffness, $B$ describes the strength of altermagnetic interaction representing anisotropic exchange interaction allowed due to lower symmetry of altermagnet, and $N_s$ is the volume saturation sublattice magnetization. The dynamics of the magnetic texture is described using the Lagrangian:
\begin{equation}\label{eq:lagr}
    L=d\times\int \frac{N_s}{\gamma}{\bf m}\cdot(\dot{\bf n}\times {\bf n}) d^2r- E,
\end{equation}
where $\gamma$ is the gyromagnetic ratio. In some cases, we include the Dzyaloshinskii-Moriya interaction (DMI). 

To drive the texture, one can use spin-transfer torques~\cite{Slonczewski1996,PhysRevB.54.9353,PhysRevB.66.224424,PhysRevLett.93.127204,PhysRevLett.106.107206,PhysRevB.80.100408}. The main result of this paper is the expression for spin-transfer torques acting on the magnetization ${\bf m}$ and the staggered field ${\bf n}$, including additional terms that account for the spin-splitter effect:
\begin{align}
    \boldsymbol{\tau}_n &= - (\bf{u}\cdot\boldsymbol{\partial})\bf{n}+\beta^\prime \bf{n} \times [{\bf u}^\prime\cdot\boldsymbol{\partial}]\bf{n} ,\label{eq:torque-n} \\
    \boldsymbol{\tau}_m &= \beta  \bf{n}\times (\bf{u}\cdot\boldsymbol{\partial})\bf{n}- [{\bf u}^\prime\cdot\boldsymbol{\partial}]\bf{n} ,\label{eq:torque-m}
\end{align}
where ${\bf u}=-\frac{g\mu_B P}{2 e N_s}{\bf j}$ is the charge drift velocity  describing the adiabatic (STT-A in Fig.~\ref{fig:sst}) spin-transfer torque in terms of some effective polarization $P$, the g-factor $g$, 
and the Bohr magneton $\mu_B$,~\cite{PhysRevB.94.054409}, $\beta$ describes the  non-adiabatic (STT-NA in Fig.~\ref{fig:sst}) contribution to the spin-transfer torque, and ${\bf j}$ is the charge current density. The terms including ${\bf u}^\prime=-\frac{g\mu_B P^\prime}{2 e N_s}(\hat{\sigma}_z\cdot{\bf j})$ describe the adiabatic (SST-A in Fig.~\ref{fig:sst}) and non-adiabatic (SST-NA in Fig.~\ref{fig:sst}) torques arising in an altermagnet due to the spin-splitting effect~\cite{PhysRevLett.126.127701} where $\hat{\sigma}_z$ is the Pauli matrix reflecting the symmetry of spin current response in altermagnet, $P^\prime$ describes the magnitude of the spin-splitting effect, and $\beta^\prime$ describes the nonadiabatic correction. The form of torques in Eqs.~\eqref{eq:torque-n} and \eqref{eq:torque-m} is established using the method in Ref.~\cite{PhysRevB.88.085423} applied to a system without spin-orbit interactions.
In particular, the first terms in Eqs.~\eqref{eq:torque-n} and \eqref{eq:torque-m} are obtained by assuming a full rotational symmetry under separate rotations of the spin space and the coordinate space, and the sublattice symmetry of ordinary antiferromagnet. The last terms are obtained by taking into account the symmetry group of a typical altermagnet with $d$-wave symmetry.

We find the Euler-Lagrange equation of motion for the slow variable $\bf{m}$, where, to leading order, 
\begin{equation}\label{eq:EL}
\begin{aligned}
{\bf m}=\frac{1}{\gamma H_{ex}}\Big [\dot{\bf n}+\Lambda{\bf n}\times(\partial_x^2{\bf n}-\partial_y^2{\bf n})\Big ]\times{\bf n} ,\\
\end{aligned}
\end{equation}
where we define $\Lambda=\gamma B/N_s$.
Using Eq.~\eqref{eq:EL}, we obtain the Lagrangian containing only the staggered field:
\begin{equation}\label{eq:lag-l}
\begin{aligned}
L=\frac{ \epsilon_0}{c^2}\int  \Big [\dot{\bf n}^2 &-c^2(\partial_\alpha {\bf\n})^2+\omega^2_0n_z^2 \\
&+{\bf \boldsymbol{\mathcal{A}}}\cdot\dot{\bf n}+\boldsymbol{\mathcal{A}}_{wz}({\bf u}^\prime\cdot\boldsymbol{\partial}){\bf{n}} \Big ]d^2r,\\
\end{aligned}
\end{equation}
with the altermagnetic interaction described by
\begin{equation}\label{eq:mag}
\begin{aligned}
\boldsymbol{\mathcal{A}}=\Lambda  {\bf n}\times(\partial_x^2 {\bf n}-\partial_y^2 {\bf n}) ,\\
\end{aligned}
\end{equation}
where $\epsilon_0= N_s d c^2/\gamma^2 H_{ex}$, $c=\gamma\sqrt{H_{ex}A/N_s}$, $\omega_0=\gamma \sqrt{H_{ex}H_{an}}$, and the time derivative should be replaced by the substitution $\dot{\bf n} \rightarrow \dot{\bf n}+({\bf u}\cdot\boldsymbol\partial) {\bf n}$ to account for the adiabatic spin-transfer torque in Eq.~\eqref{eq:torque-n}.
The adiabatic spin-splitter torque is included in the Lagrangian \eqref{eq:lag-l} by adding a term containing the vector potential of the Wess-Zumino action $\boldsymbol{\mathcal{A}}_{wz}$ ~\cite{PhysRev.74.817,PhysRevB.98.224401,PhysRevB.81.014414,Kovalev2012}, i.e., $\nabla_{\bf n}\times \boldsymbol{\mathcal{A}}_{wz}=2\gamma H_{ex}{\bf n}$. The non-adiabatic torques in Eqs.~\eqref{eq:torque-n} and \eqref{eq:torque-m} can be included via the Rayleigh function:
\begin{equation}\label{eq:RF}
\begin{aligned}
\mathcal{R}=\frac{\epsilon_0 \gamma H_{ex}}{c^2}\left[\alpha \dot{\bf n}^2+ 2\beta\dot{\bf n} \cdot({\bf u}\cdot\boldsymbol\partial) {\bf n}+2\beta^\prime\dot{\bf m} \cdot({\bf u}^\prime\cdot\boldsymbol\partial) {\bf n}\right],\\
\end{aligned}
\end{equation}
where $\alpha$ is the Gilbert damping parameter. Note that the Euler-Lagrange-Rayleigh equation derived from Eqs.~\eqref{eq:lagr} and \eqref{eq:RF} yields an additional higher-order term, $\boldsymbol \tau_m = \beta^\prime \mathbf{m} \times [\mathbf{u}^\prime \cdot \boldsymbol{\partial}] \mathbf{n}$, in Eq.~\eqref{eq:torque-m}, which is also allowed by symmetry. In Eq.~\eqref{eq:lag-l}, we switch to dimensionless variables arriving at
\begin{equation}\label{eq:dim}
\begin{aligned}
L_0= \int  \Big [\dot{\bf n}^2 -(\partial_\alpha {\bf\n})^2+n_z^2+{\bf \boldsymbol{\mathcal{A}}}\cdot\dot{\bf n}+\boldsymbol{\mathcal{A}}_{wz}({\bf u}^\prime_0\cdot\boldsymbol{\partial}){\bf{n}} \Big ]d^2r.\\
\end{aligned}
\end{equation}
Here $\dot{\bf n} \rightarrow \dot{\bf n}+({\bf u}/c\cdot\boldsymbol\partial) {\bf n}$, the unit of speed is $c$, that of time $1/\omega_0$, and that of length $c/\omega_0$, $\nabla_{\bf n}\times \boldsymbol{\mathcal{A}}_{wz}=2{\bf n}$. We introduce the dimensionless ${\bf u}^\prime_0=(\gamma H_{ex}/\omega_0){\bf u}^\prime/c$, ${\bf u}_0= (\gamma H_{ex}/\omega_0){\bf u}/c$, and  notation $\alpha_0=(\gamma H_{ex}/\omega_0)\alpha$, arriving at dimensionless expression for the Rayleigh function: 
\begin{equation}\label{Eeq:RF1}
\begin{aligned}
\mathcal{R}_0=\alpha_0 \dot{\bf n}^2+ 2\beta\dot{\bf n} \cdot({\bf u}_0\cdot\boldsymbol\partial) {\bf n}+2\beta^\prime\dot{\bf m} \cdot({\bf u}^\prime_0\cdot\boldsymbol\partial) {\bf n}\,.\\
\end{aligned}
\end{equation}
Dimensionless altermagnetic interaction is $\Lambda_0=\Lambda \omega_0/c$. 
\begin{figure}
    \centering
    \includegraphics[width=\columnwidth]{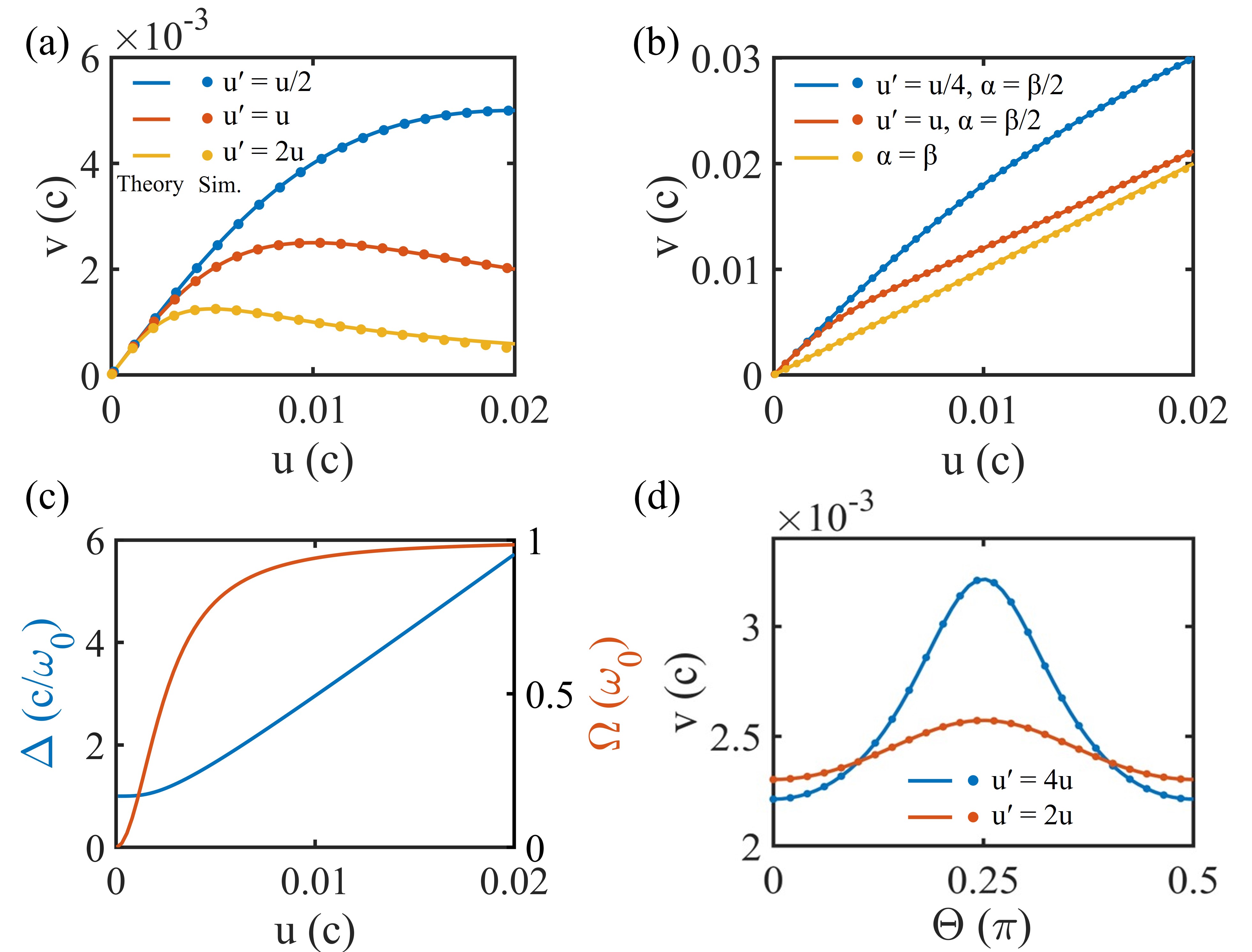}
    \caption{Current-induced dynamics of a domain wall. (a) The domain wall velocity as a function of the charge drift velocity \( u \), with the added spin-splitter torque described by \(u'\) and the adiabatic spin-transfer torque turned off. (b) The same, but in the presence of the adiabatic spin-transfer torque. (c) The angular precession speed \( \Omega \) and the domain wall width \( \Delta \) as functions of the charge drift velocity, with \( u = 2u' \). (d) The domain wall velocity for different directions of the charge current with respect to crystallographic axes described by \( \Theta \), with \( u = 0.0013c \). Parameters used:  $\beta = 0.01, \quad H_{\text{an}} = 1.45 \, \mathrm{T}, \quad N_s = 10 \, \mathrm{kA/m}, \quad c = 35 \, \mathrm{km/s}, \quad \beta' = \beta, \quad \alpha = \beta, \text{ except in (b)}$.}
    \label{fig:dw_dynamic2}
\end{figure}

To perform micromagnetics of a thin layer of altermagnet, we use mumax3 code~\cite{Vansteenkiste2014} implementing two AFM coupled ferromagnetic layers, with anisotropic intralayer exchange interaction described by exchange stiffness parameters $A^{ex}_x$ and $A^{ex}_y$~\cite{PhysRevLett.133.196701}. For micromagnetics, we use parameters corresponding to Ref.~\cite{Gomonay2024}: the lattice spacing $a_0=0.448$nm, the intralayer exchange $A=A^{ex}_x+A^{ex}_y=2.7\times10^{-12}$J/m, the AFM coupling between layers $a_0^2 H_{ex} N_s/2=4A^{ex}_z=3.2\times10^{-11}$J/m, and the altermagnetic interaction corresponding to anisotropy in the exchange $B=2(A^{ex}_x-A^{ex}_y)=1.1\times10^{-12}$J/m, where the exchange anisotropy has opposite sign in the second layer, see Supplemental Material for details~\cite{Note}.

\textit{Domain wall dynamics.} -- 
It has been found that altermagnetic interaction can affect the Walker breakdown and oscillatory dynamics of magnetic domain walls in altermagnets~\cite{Gomonay2024}. Here, we explore how the domain wall dynamics is affected by the spin-splitter torque in a magnetic wire. We describe the staggered field in spherical coordinates, i.e., ${\bf n}=(\sin \theta \cos \phi,\cos \theta \sin \phi, \cos \theta)$. We use the collective coordinate approach with the following ansatz that describes the domain wall profile~\cite{Kosevich1990,Gomonay2024}:
\begin{gather}\label{eq:ansatz-dw}
    \cos\theta(x^\prime,t)=-p\,\tanh\frac{x^\prime-X(t)}{\Delta(t)}, \\
    \phi(x^\prime,t)=\Phi(t)+b(t)\frac{x^\prime-X(t)}{\Delta(t)},
\end{gather}
where $p=\pm 1$ is the topological charge, $X(t)$ and $\Phi(t)$ are the collective variables that describe the position and the tilt along the direction of the wire defined by $\hat{x}^\prime$, and the collective variables $b(t)$ and $\Delta(t)$ are generally time dependent but can be treated as slow variables. After writing the equations of motion for the collective variables, we can find the slow variables $b$ and $\Delta$. In particular, the width of the domain wall is given by the expression $\Delta=\Delta_0[1-(v-u)^2/c^2]/\sqrt{1-(v-u)^2/c^2-\Omega^2/\omega_0^2}$ where $\Delta_0=c/\omega_0$ is the width of the equilibrium domain wall. With substituted $b$ and $\Delta$, we obtain the Lagrangian expressed in terms of $X(t)$ and $\Phi(t)$:
\begin{align}
    L&=-4 \sqrt{1-(\dot{X}-u_0)^2-\dot{\Phi}^2} +2\Phi \tilde{u}_0^\prime,
\end{align}
where $\tilde{u}_0^\prime=({\bf u}^\prime_0 \cdot {\bf u}_0)/u_0$ and we assume that the charge current is along the direction of the wire, i.e., ${\bf u}=u(\cos\Theta,\sin\Theta)$.  After including the Rayleigh function, we solve the Euler-Lagrange equations for a steady stationary state for which $X(t)=vt$ and $\Phi(t)=\Omega t$, and obtain the following expressions for the domain wall velocity and the angular speed of precession:
\begin{align} \label{eq:an1}
    v=&\frac{\tilde{u}^{\prime2} +(\beta^2-\alpha^2)u^2+\alpha^2c^2}{2 \alpha (\beta-\alpha)  u} \\ \nonumber 
    &-\sqrt{\left(\frac{\tilde{u}^{\prime2} +(\beta-\alpha)^2u^2+\alpha^2c^2}{2 \alpha (\beta-\alpha)  u}\right)^2-c^2},\\ \label{eq:an2}
    \frac{\Omega}{\omega_0}=&\frac{1}{2}-\frac{\tilde{u}^{\prime2} +\alpha^2c^2}{2 (\beta-\alpha)^2  u^2} \\  
    &+\sqrt{\left(\frac{\tilde{u}^{\prime2} +(\beta-\alpha)^2u^2+\alpha^2c^2}{2  (\beta-\alpha)^2  u^2}\right)^2-\frac{c^2\alpha^2}{(\beta-\alpha)^2  u^2}},\nonumber
\end{align}
where $\tilde{u}^\prime=({\bf u}^\prime \cdot {\bf u})/u=(\cos\Theta^2-\sin\Theta^2)u$. Note that both expressions have a well-defined limit for $\beta\rightarrow \alpha$, as in this limit we obtain $v \rightarrow u$. To obtain the above results, we dropped the terms containing the product $\Lambda_0 \beta^\prime$ associated with the nonadiabatic spin-splitter torque because their contribution is small as long as $\Lambda_0 \beta^\prime \ll1$. We observe good agreement of the above result with micromagnetics for the domain wall velocities $v < c$.
\begin{figure}
    \centering
    \includegraphics[width=\columnwidth]{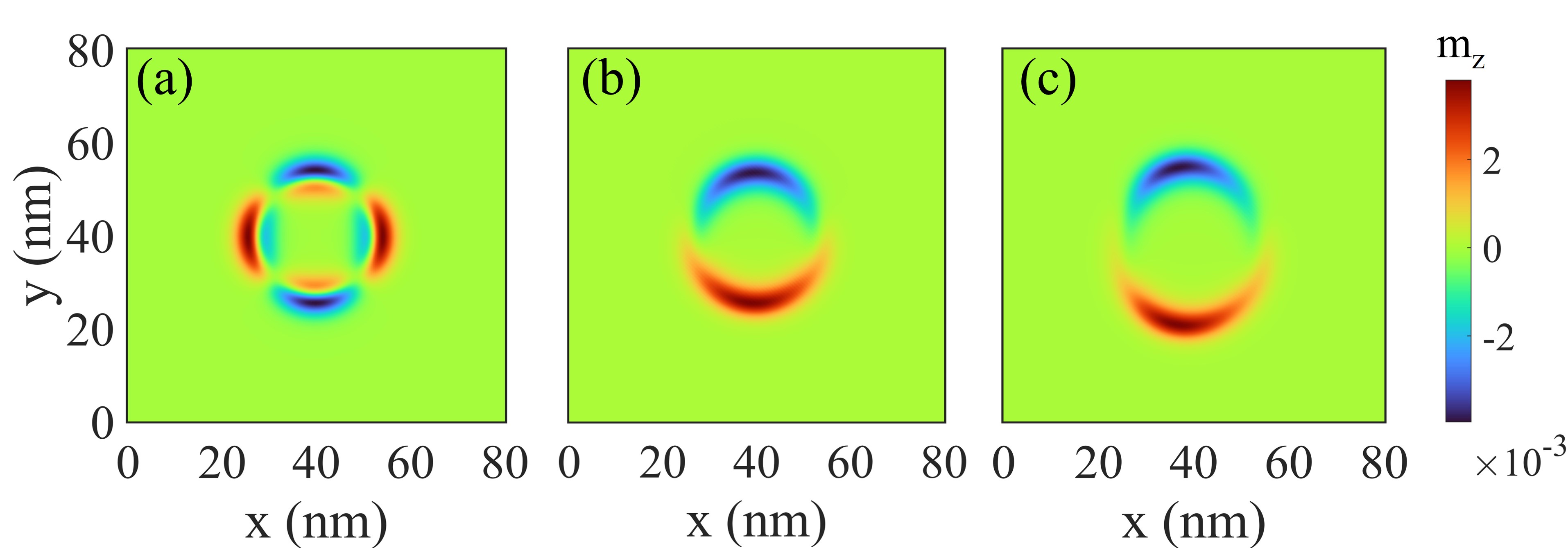}
    \caption{Out-of-plane magnetization of: (a) A stationary skyrmion, and (b) and (c) moving skyrmions in the presence of spin-transfer and spin-splitter torques, described by $u = u^\prime = 50$ m/s and $u = u^\prime = 100$ m/s, respectively. The charge current is applied in the $y$-direction. Parameters used: $\alpha = 0.01$, $\beta = 0.01$, $H_{an} = 1.45$ T, $N_s = 10~kA/m$, $D = 0.55 $ mJ/m$^2$.}
    \label{fig:Mumax3_demag}
\end{figure}

In Fig.~\ref{fig:dw_dynamic2}, we plot the results of Eqs.~\eqref{eq:an1} and \eqref{eq:an2} along with the results of the micromagnetic simulations. To clearly identify the effect of domain wall precession induced by the adiabatic spin-splitter torque, in Fig.~\ref{fig:dw_dynamic2}(a) we switch off the adiabatic spin-transfer torque in Eq.~\eqref{eq:torque-n}. We observe a typical behavior identified before in Ref.~\cite{Belashchenko2016} for magnetoelectric antiferromagnets where the precession hinders the domain wall motion leading to the appearance of maximally attainable velocity. In Fig.~\ref{fig:dw_dynamic2}(b), we include the adiabatic spin-transfer torque and observe that it can suppress the precession of the domain wall in the limit when $\beta \rightarrow \alpha$. This is consistent with the fact that when $\alpha=\beta$ we obtain $v=u$.
However, when $\alpha \ne \beta$ the effect of precession will result in a non-linear response to the applied current, as can be seen in Fig.~\ref{fig:dw_dynamic2}(b). For further insight, in Fig.~\ref{fig:dw_dynamic2}(c) we plot the angular frequency of domain wall precession and the domain wall width as a function of $u$. For domain wall precession, we observe saturation to the maximally attainable angular frequency $\omega_0$ while the domain wall width increases which is in contrast to the Lorentz contraction of domain walls in antiferromagnets.

Finally, in Fig.~\ref{fig:dw_dynamic2}(d) we study the anisotropy of domain wall speed for different current directions with respect to the crystallographic axes of altermagnet described by $\Theta$. For the direction aligned with the large spin-splitter effect for which 
 $|\tilde{u}^\prime|=u$, we observe the slowest speed of domain wall, which can be explained by the domain wall precession induced by the spin-splitter torque.

\textit{Skyrmion dynamics.} --
To stabilize skyrmion, we assume the presence of interfacial DMI in Eq.~\eqref{eq:hamiltonian},
\begin{equation}
    E_{dmi}=d \times\int D [n_z \boldsymbol{\partial}\cdot {\bf n}- ({\bf n}\cdot\boldsymbol{\partial}) n_z] d^2r,
\end{equation}
where $D$ describes the strength of DMI. In dimensionless units, the DMI strength is described by $D_0=D c/A_0 \omega_0$.  We assume a traveling wave solution of a form ${\bf n}({\bf r}-{\bf v} t)$. After substituting a traveling wave form in Eq.~\eqref{eq:dim}, we can find the shape of skyrmion by extremizing a time-independent Lagrangian numerically.
For speeds that are much smaller than $c$, the solutions are close to the $360^{\circ}$ domain wall ansatz~\cite{Bttner2018,Wang2018,PhysRevB.50.16485}, while at large speeds we recover the "relativistic`` deformation of the skyrmion~\cite{PhysRevB.106.L220402,SciPostPhys.8.6.086}. 
This deformation can be seen in Fig.~\ref{fig:Mumax3_demag} where we plot the net magnetization distribution. The net magnetization distribution of a moving skyrmion is distinct from that of a static skyrmion, where in the latter case the net magnetization arises exclusively due to altermagnetic interaction.

\begin{figure}
    \centering
    \includegraphics[width=\columnwidth]{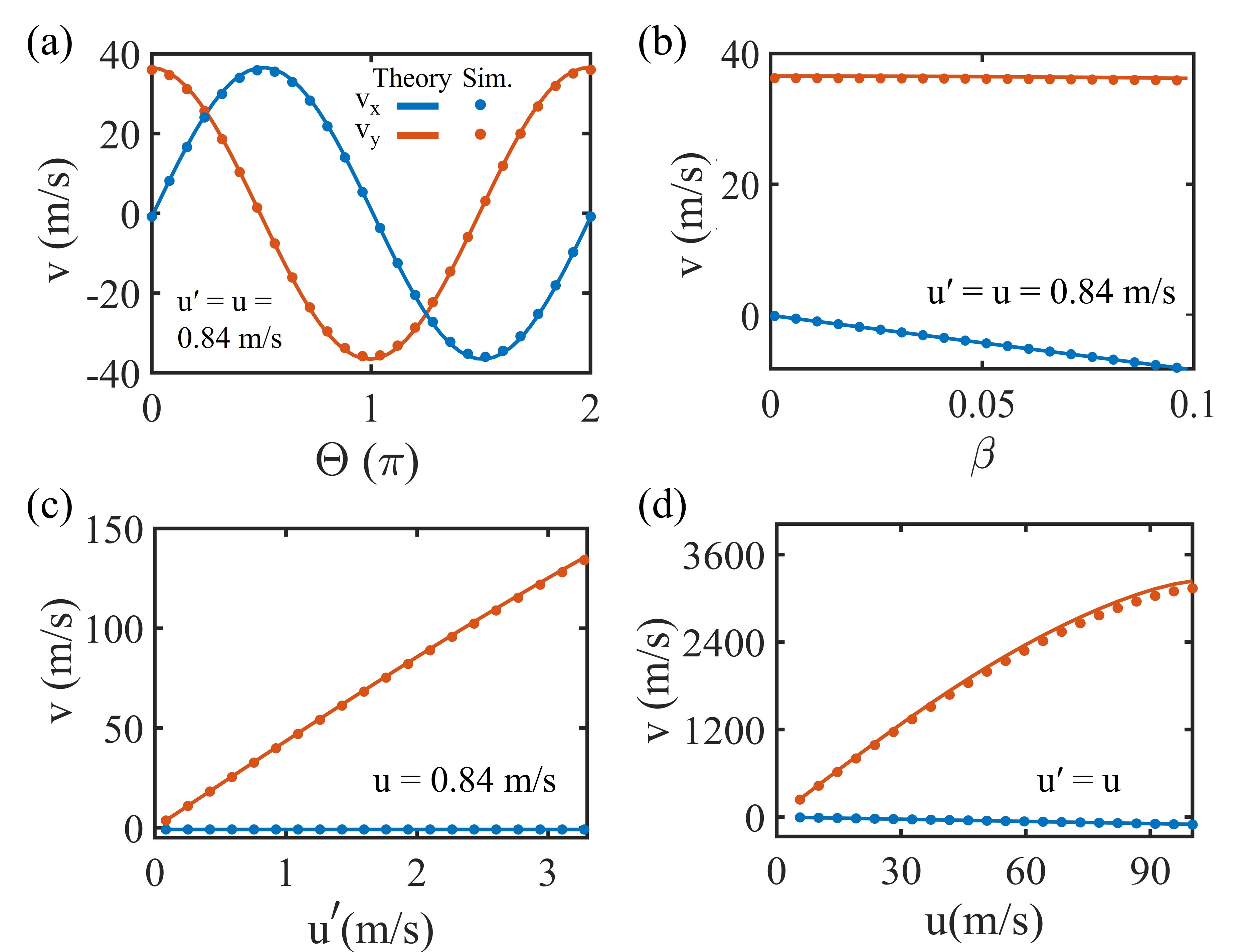}
    \caption{Components of the skyrmion velocity, $v_x$ and $v_y$, calculated from micromagnetics and from Eq.~\eqref{eq:velocity} for different directions of the charge current with respect to crystallographic axes described by $\Theta$. The spin-transfer torque is parametrized by \(u\), and the spin-splitter torque by \(u'\). Parameters used: $\alpha = 0.01$, $\beta = 0.01$ (in (a), (c), and (d)), $H_{an} = 1.45$ T, $N_s = 10~kA/m$, $D = 0.55$ mJ/m$^2$. The current is applied in the $x$-direction ($\Theta = 0 $) in (b)-(d).   }
    \label{fig:velocity}
\end{figure}
We further parametrize the skyrmion position in terms of the generalized coordinate ${\bf X}(t)$ and use the Thiele equation to describe the skyrmion dynamics analytically. The Lagrangian \eqref{eq:lag-l} leads to the equation for ${\bf X}(t)$: 
\begin{equation}\label{eq:sk-dynamics}
    \hat{\mathcal{M}} \Ddot{\bf X}+\hat{\mathcal{G}}  \dot{\bf X} +\alpha_0 \hat{\mathcal{D}} \dot{\bf X}+\hat{\mathcal{B}}(\dot{\bf X}-{\bf u}/c)={\bf F},
\end{equation}
where $\mathcal{\hat{M}}$ is the skyrmion mass tensor, $\mathcal{\hat{G}}$ is the antisymmetric gyrotensor, $\mathcal{D}_{ij}=\int {\partial_i {\bf n}\cdot \partial_j {\bf n}} d^2 r$ is the dissipative tensor, and $\mathcal{B}_{ij}=\int [\partial_i(\boldsymbol{\mathcal{A}}\cdot\partial_j {\bf n})-\partial_j(\boldsymbol{\mathcal{A}}\cdot\partial_i {\bf n})] d^2 r$. Our Eq.~\eqref{eq:sk-dynamics} differs from the result obtained in Ref.~\cite{PhysRevLett.133.196701}, see \cite{Note} for details. For a compensated antiferromagnet the gyrotensor vanishes as the total topological charge of skyrmion is zero; however, the weak ferromagnetism can result in some nonvanishing gyrotensor. Here, we consider a compensated antiferromagnet, and disregard the effect of the nonadiabatic spin-splitter torque in Eq.~\eqref{Eeq:RF} by setting $\beta^\prime=0$, as its effect is small. For the force, we obtain the following expression: 
\begin{equation}
    {\bf F}=4\pi\hat{z}\times {\bf u}^\prime_0+\beta \mathcal{\hat{D}}\cdot {\bf u}_0.
\end{equation}
Due to the presence of the spin-splitter effect the direction of the force is not aligned with the current direction, and we expect the skyrmion Hall effect induced by the spin-splitter torque. For the velocity we obtain:
\begin{align}\label{eq:velocity-long}
v_x=\frac{(b^2+b\eta+\alpha\beta|\hat{\mathcal{D}}|)u_x-bu_y^+(\alpha-\beta)+\alpha\eta u_y^-}{b^2+|\hat{\mathcal{D}}|\alpha^2}, \\
v_y=\frac{(b^2-b\eta+\alpha\beta|\hat{\mathcal{D}}|)u_y+bu_x^+(\alpha-\beta)+\alpha\eta u_x^-}{b^2+|\hat{\mathcal{D}}|\alpha^2},
\end{align}
where $|\hat{\mathcal{D}}|=\det\hat{\mathcal{D}}$, ${\bf u}^\pm=\hat{\mathcal{D}}^\pm{\bf u}$, $\hat{\mathcal{D}}^\pm=((\mathcal{D}_{xx},\pm\mathcal{D}_{xy}),(\pm\mathcal{D}_{xy},\mathcal{D}_{yy}))$, $\eta=4 \pi P^\prime/P$, and $b=\mathcal{B}_{xy}/(\gamma H_{ex}/\omega_0)$. 
We expect the effect of the altermagnetic interaction described by $b$ to be generally small, and to vanish exactly for a rotationally symmetric skyrmion. We confirmed this by micromagnetic simulations.
By setting $b=0$, we obtain the simplified expressions: 
\begin{equation}\label{eq:velocity}
v_x=\frac{u_x\beta|\hat{\mathcal{D}}|+u_y^-\eta}{|\hat{\mathcal{D}}|\alpha},\quad
v_y=\frac{u_y\beta|\hat{\mathcal{D}}|+u_x^-\eta}{|\hat{\mathcal{D}}|\alpha}.
\end{equation}

In Fig. ~\ref{fig:velocity}, we plot the results of Eq.~\eqref{eq:velocity} along with the results of the micromagnetic simulations.
In Fig.~\ref{fig:velocity}(a), the components of velocity for different current directions exhibit the skyrmion Hall effect with the expected $d$-wave symmetry. In Fig.~\ref{fig:velocity}(b), we study how the nonadiabatic parameter $\beta$ affects the skyrmion velocity. We observe that only the longitudinal motion is affected substantially. In contrast, the spin-splitter adiabatic torque can substantially affect the transverse component of the velocity due to the Magnus force, as can be seen in Fig.~\ref{fig:velocity}(c). As shown in Fig.~\ref{fig:velocity}(d), the spin-splitter adiabatic torque can lead to very fast skyrmion motion, which is typically not possible in antiferromagnets with the spin-transfer torque alone when $\alpha$ is comparable to $\beta$.

 \textit{Conclusions.} --
 We have identified the adiabatic and nonadiabatic spin-splitter torques in textured $d$-wave altermagnets. Using the Lagrangian formalism, we have incorporated these torques into the description of magnetic textures, and our results can be used for micromagnetic studies of altermagnets. We have found that the spin-splitter torque can substantially modify the behavior of magnetic domain walls by slowing down the domain wall motion through the effect of domain wall precession. Thus, altermagnets exhibit certain properties of ferromagnets and ferrimagnets while differing from antiferromagnets, in which current-induced domain wall precession is absent. In the case of magnetic skyrmions, the spin-splitter torque induces an anisotropic skyrmion Hall effect, which is absent in antiferromagnets but can be present in ferromagnets and ferrimagnets. Surprisingly, much faster current-induced motion of skyrmions is possible in the presence of spin-splitter torque, in contrast to antiferromagnetic skyrmions driven by nonadiabatic torque. Similar physics arises for skyrmions in ferromagnets due to the Magnus force combined with the effect of low Gilbert damping.~\cite{Sampaio2013,Ohki2024}. What differentiates altermagnets from other magnetic systems is that the dynamics of the magnetic textures exhibit characteristic anisotropy reflecting the $d$-wave symmetry. Our findings can serve as a hallmark of altermagnetism in textured magnets and pave the way for spintronics applications.

\textit{Acknowledgments.} --
We acknowledge useful discussions with K. Belashchenko.
This work was supported by the U.S. Department of Energy, Office of Science, Basic Energy Sciences, under Award No. DE-SC0021019.

\textit{Note added:} We became aware of another manuscript appearing on arXiv after our submission, in which a similar expression for the spin-splitter adiabatic torque is proposed~\cite{zarzuela2024}. These ideas can be further extended to superconducting systems~\cite{kokkeler2025}.

\bibliography{lib}
\widetext
\clearpage
\begin{center}
    {\large Supplemental Material for}\vspace{5mm}\\
    \textbf{\large Spin-transfer torque in altermagnets with magnetic textures}
\end{center}
\begin{center}
    \begin{normalsize}
        {Hamed Vakili, Edward Schwartz, Alexey A. Kovalev}\\
        \vspace{1mm}
        \textit{Department of Physics and Astronomy and Nebraska Center for Materials and Nanoscience,\\ University of Nebraska, Lincoln, Nebraska 68588, USA}
    \end{normalsize}
\end{center}
\section*{Model}  

We have used a two sublattice system with the Hamiltonian described as:
\begin{align}\label{eq:ham-alt}
    H_{AM} &= -\sum_{i,j}\Bigl[ J_x \boldsymbol{s}_{i,j}^A.\boldsymbol{s}_{i\pm 1,j}^A + J_y \boldsymbol{s}_{i,j}^A.\boldsymbol{s}_{i,j\pm1}^A \nonumber\\ 
&+ J_y \boldsymbol{s}_{i,j}^B.\boldsymbol{s}_{i\pm 1,j}^B + J_x \boldsymbol{s}_{i,j}^B.\boldsymbol{s}_{i,j\pm 1}^B+ 2J_3 \boldsymbol{s}_{i,j}^A.\boldsymbol{s}_{i,j}^B\nonumber\\
&+ K_0 (\boldsymbol{s}_{i,j}^A.z)^2 + K_0 (\boldsymbol{s}_{i,j}^B.z)^2\Bigr] ~.
\end{align}
$J_x, J_y$ are the intralayer ferromagnetic coupling, $J_3$ is the interlayer antiferromagnetic coupling and $K_0$ is the uniaxial anisotropy. $\boldsymbol{s_{i,j}^A}, \boldsymbol{s_{i,j}^B}$ are the onsite normalized spins. Their relationship to micromagnetic parameters in the main text are as follows:
\begin{align}
    A_x^{ex} = J_x/a_0, A_y^{ex} = J_y/a_0, A_z^{ex} = J_3/a_0, H_{an}N_s = 2K_0/a^3~,
\end{align}
where $a_0$ is the lattice constant. The volume saturation sublattice magnetization ($N_s$) is related by $N_s = \mu/a_0^3$ to the onsite spins, where $\mu$ is the magnetic moment of each site. The simulation parameters are presented in the table below:
\begin{table}[h!]
    \centering
    \begin{tabular}{l|l||l|l}

    \multicolumn{2}{c||}{\textbf{Atomistic parameters}} & \multicolumn{2}{c}{\textbf{Micromagnetic parameters}} \\ \hline
    Lattice constant         & $a_0 = 0.448$ nm    & Intralayer exchange        & $A = 2.7$ pJ/m        \\
    Magnetic moment           & $\mu = \mu_B$    & Saturation magnetization  & $N_s = 10$ kA/m     \\
    Inter-layer AFM exchange & $J_3 = 22.4$ meV   & AFM coupling & $A_{z}^{ex} = 8$ pJ/m       \\
    FM exchange -x direction  & $J_x = 4.54 $ meV   & FM stiffness -x direction  & $A^{ex}_x = 1.625$ pJ/m      \\
    FM exchange -y direction     & $J_y = 3.0$ meV     & FM stiffness -y direction   & $A^{ex}_y = 1.075$ pJ/m         \\
    Uniaxial anisotropy     & $K_0 = 0.042$ meV     & Anisotropy field   & $H_{an} = 1.45$ T         \\
    \hline
    \end{tabular}
    \caption{Atomistic and their corresponding micromagnetic parameters used in this paper. }
    \label{tab:my_label}
\end{table}

Note that Eq.~\eqref{eq:ham-alt} corresponds to the simplest possible representation of an altermagnet, describing what is usually referred to as the long-wavelength or micromagnetic limit. A more accurate description of the magnetic sublattices of an altermagnet (e.g., going beyond AA stacking~\cite{Gomonay2024}) is possible within mumax3~\cite{Vansteenkiste2014} using functionality such as the AddFieldTerm() and Shifted() functions. We have implemented this approach using the magnetic lattice of prototypical RuO$_2$ and did not observe any differences within the parameter range used in our paper. This confirms that our simulations are safely within the regime where micromagnetics is fully adequate.

\noindent
\section*{Micromagnetic implementation of spin-splitter torque} 

To implement Eqs.~(3) and (4) of the main text, we used the Zhang-Li torque functionality in Mumax3~\cite{Vansteenkiste2014}. By choosing different currents for each sublattice (implemented as separate layers), we obtain torques:
\begin{align}
\boldsymbol \tau_1 &= - ({\bf u}_1\cdot\boldsymbol{\partial}){\bf m}_1+\beta_1 {\bf m}_1 \times [{\bf u}_1\cdot\boldsymbol{\partial}]{\bf m}_1 ,\label{eq:torque1} \\
\boldsymbol \tau_2 &= - ({\bf u}_2\cdot\boldsymbol{\partial}){\bf m}_2+\beta_2 {\bf m}_2 \times [{\bf u}_2\cdot\boldsymbol{\partial}]{\bf m}_2 ,\label{eq:torque2}
\end{align}
where the parameters ${\bf u}_1$, ${\bf u}_2$, $\beta_1$, and $\beta_2$ can be chosen arbitrarily for the two sublattices.
This leads to implementation of torque in Eqs.~(3) and (4) of the main text with ${\bf u}=({\bf u}_1+{\bf u}_2)/2$, ${\bf u}^\prime=({\bf u}_1-{\bf u}_2)/2$, and $\beta=\beta^\prime=\beta_1=\beta_2$. By properly choosing ${\bf u}_1$ and ${\bf u}_2$ we can realize an arbitrary direction of ${\bf u}$ and ${\bf u}^\prime$ that satisfies the constraint ${\bf u}^\prime=\frac{P^\prime}{P}\sigma_z\cdot{\bf u}$. In this implementation, we always have $\beta=\beta^\prime$. 

\section*{Dynamics of skyrmion induced by spin-transfer torque} 
Here we consider the dynamics of skyrmion in the presence of only spin-transfer torque, i.e., when ${\bf u}^\prime=0$.
We parametrize the skyrmion position in terms of the generalized coordinate ${\bf X}(t)$ and use the Thiele equation to describe the skyrmion dynamics analytically. This leads to the equation for ${\bf X}(t)$: 
\begin{equation}\label{eq:sk-dynamics}
    \hat{\mathcal{M}} \Ddot{\bf X}+\hat{\mathcal{G}}  \dot{\bf X} +\alpha_0 \hat{\mathcal{D}} \dot{\bf X}+\hat{\mathcal{B}}(\dot{\bf X}-{\bf u}/c)=\beta \mathcal{\hat{D}}\cdot {\bf u}_0,
\end{equation}
where $\mathcal{\hat{M}}$ is the skyrmion mass tensor, $\mathcal{\hat{G}}$ is the antisymmetric gyrotensor, $\mathcal{D}_{ij}=\int {\partial_i {\bf n}\cdot \partial_j {\bf n}} d^2 r$ is the dissipative tensor, $\mathcal{B}_{ij}=\int [\partial_i(\boldsymbol{\mathcal{A}}\cdot\partial_j {\bf n})-\partial_j(\boldsymbol{\mathcal{A}}\cdot\partial_i {\bf n})] d^2 r$ describes altermagnetic interaction, ${\bf u}_0= (\gamma H_{ex}/\omega_0){\bf u}/c$, and  $\alpha_0=(\gamma H_{ex}/\omega_0)\alpha$. For a compensated antiferromagnet the gyrotensor vanishes as the total topological charge of skyrmion is zero. 

Equation~\eqref{eq:sk-dynamics} leads to the following components of the velocity:
\begin{align}\label{eq:velocity-long}
v_x=\frac{(b^2+\alpha\beta|\hat{\mathcal{D}}|)u_x-bu_y^+(\alpha-\beta)}{b^2+|\hat{\mathcal{D}}|\alpha^2}, \\
v_y=\frac{(b^2+\alpha\beta|\hat{\mathcal{D}}|)u_y+bu_x^+(\alpha-\beta)}{b^2+|\hat{\mathcal{D}}|\alpha^2},\label{eq:velocity-long1}
\end{align}
where $|\hat{\mathcal{D}}|=\det\hat{\mathcal{D}}$, ${\bf u}^\pm=\hat{\mathcal{D}}^\pm{\bf u}$, $\hat{\mathcal{D}}^\pm=((\mathcal{D}_{xx},\pm\mathcal{D}_{xy}),(\pm\mathcal{D}_{xy},\mathcal{D}_{yy}))$, and $b=\mathcal{B}_{xy}/(\gamma H_{ex}/\omega_0)$. 
Assuming an elliptical skyrmion profile, it can be shown that $\mathcal{B}_{xy}$ scales with the square of the eccentricity in the lowest-order approximation. Therefore, we expect the effect of the altermagnetic interaction described by $b$ to be generally small, and to vanish exactly for a rotationally symmetric skyrmion. We also confirmed this by micromagnetic simulations. This shows that for chosen parameters, the altermagnetic interaction can only induce a negligibly small skyrmion Hall effect.

We note that Eqs.~\eqref{eq:sk-dynamics}, \eqref{eq:velocity-long}, \eqref{eq:velocity-long1} substantially differ from similar equations presented in Ref.~\cite{PhysRevLett.133.196701}. Our results correctly reproduce the limit $\alpha = \beta$, where, based on the general form of the Lagrangian and the Rayleigh function, it is expected that the soliton moves with the electron drift velocity ${\bf u}$. Equations~(10), (11), (12) in Ref.~\cite{PhysRevLett.133.196701} do not correctly reproduce the limit $\alpha = \beta$.

\end{document}